\documentclass{emulateapj}

\newcommand{\mincir}{\raise
-3.truept\hbox{\rlap{\hbox{$\sim$}}\raise4.truept\hbox{$<$}\ }}
\newcommand{\magcir}{\raise
-3.truept\hbox{\rlap{\hbox{$\sim$}}\raise4.truept\hbox{$>$}\ }}

\makeatother

\shorttitle{The Local Environment of Bright IRAS Galaxies: The AGN/Starburst connection.}
\shortauthors{Koulouridis et al.}

\begin{document}

\title{A 3-Dimensional study of the Local Environment of Bright IRAS
  Galaxies: The AGN/Starburst connection.}

\author{Elias Koulouridis\altaffilmark{1,3},  Vahram Chavushyan\altaffilmark{2,4}, 
Manolis Plionis\altaffilmark{1,2}, Yair Krongold\altaffilmark{4},  and Deborah Dultzin-Hacyan\altaffilmark{4}}

\altaffiltext{1}{Institute of Astronomy \& Astrophysics, National Observatory of
Athens, I. Metaxa \& B. Pavlou, P. Penteli 152 36, Athens, Greece}

\altaffiltext{2}{Instituto Nacional de Astrof\'{\i}sica Optica y Electr\'onica,
A.P. 51 y 216, C.P. 72000, Puebla, Pue., M\'exico}

\altaffiltext{3}{Physics Department, Univ. of Patras, Panepistimioupolis Patron, 26500, Patras, Greece}

\altaffiltext{4}{Instituto de Astronom\'ia, Univesidad Nacional
Aut\'onoma de M\'exico, A.P. 70-264, M\'exico, D. F.
04510, M\'exico}

\begin{abstract}
We present a 3-dimensional study of the local  ($\leq 100 \; h^{-1}$
kpc) and the large scale ($\leq$ 1 $h^{-1}$ Mpc) 
environment of Bright IRAS Galaxies (BIRGs).
For this purpose we use 87 BIRGs located at
high galactic latitudes (with 0.008$\leq z \leq$0.018) as well as a control sample of non-active
galaxies having the same morphological, redshift and diameter size
distributions as the corresponding BIRG sample.
Using the Center for Astrophysics (CfA2) and 
Southern Sky Redshift Survey (SSRS) galaxy catalogues ($m_b\lesssim 15.5$)
as well as our own spectroscopic observations ($m_b\lesssim19.0$) for a
subsample of the original BIRG sample, we find that the fraction of BIRGs with a
close neighbor is significantly higher than that of their control sample.
Comparing with a related analysis of Sy1 and Sy2 galaxies of
Koulouridis et al. (2006) we find that BIRGs have a similar environment as
Sy2s, although the fraction of BIRGs with a bright close neighbor is even
higher than that of Sy2 galaxies.
An additional analysis of the relation between FIR colors and the type
of activity of each BIRG shows a significant difference between 
the colors of strongly-interacting and non-interacting starbursts 
and a resemblance between the colors of non-interacting starbursts and Sy2s.
Our results support the view where close interactions 
can drive molecular clouds towards the galactic
center, triggering starburst activity and obscuring the nuclear 
activity. When the close neighbor moves away, starburst activity is reduced
with the simultaneous appearance of an obscured (type 2) AGN. 
Finally, the complete disentanglement of the pair gives 
birth to an unobscured (type 1) AGN.

\end{abstract}


\section{Introduction}

The IRAS Revised Bright galaxy sample by Sanders et al. (2003)
includes all galaxies with total 60 $\mu m$ flux density greater than 5.24 Jy. 
The sample is the result of a highly complete flux-limited survey
conducted by IRAS covering the entire sky at galactic latitudes $|b|\geq 5^o$
and was  compiled after the final calibration of the IRAS Level 1 Archive. 
It offers far more accurate and consistent measurements of the 
flux of objects with extended emission. In addition, the 
infrared fluxes of over 100 sources from the sample were 
recalculated by the IRAS High Resolution (HIRES) processing, 
which allowed the deconvolution of close galaxy pairs 
(Surace, Sanders, Mazzarela 2004). The latter provides a 
more-than-ever reliable database of the IRAS galaxies 
which can be proved crucial for statistical studies like this one.

While the relation between Ultra Luminous IRAS Galaxies 
(ULIRGs) and strong interactions has been thoroughly studied 
(e.g. Sanders, Surace \& Ishida 1999, Wang et al. 2006), this is not the case for the 
environment of moderately and low luminous infrared 
galaxies. A 2-dimensional analysis of Krongold et al. (2002)
showed a trend for a Bright IRAS Galaxy (BIRG) sample on having neighbors
in excess of normal galaxies and Sy1 galaxies, but 
in relative agreement with Sy2 galaxies. However,
the BIRG population consists of various types of active galaxies, 
including starbursts (the majority), Seyferts, Liners and 
normal galaxies and thus it would be of great interest to 
clarify the connection between infrared emission, interactions 
and different types of active galaxies.

During the last decade, many studies have investigated the 
relation among interacting galaxies, starbursting and 
nuclear activity (eg. Hernandez-Toledo et al. 2001;  Ho 2005).
Despite the plethora 
of available information, searches for correlations 
between the above physical processes
are inconclusive, the only exception being the coupling 
between interactions and starbursting. However, there is 
evidence that AGN galaxies host a post-starburst stellar 
population (eg. Boisson et al. 2000, Gonz\'alez Delgado et al. 2001)
while Kauffmann et al. (2003) showed that the
 fraction of post-starburst stars increases with AGN emission. 
Proving a relation of this type between starburst and AGN galaxies
would simultaneously solve also the problem of the AGN triggering
mechanism. 
Interactions would be the main cause of such activities, being
starbursting and/or the feeding of a central black hole. 
However, this is not a trivial task. The main difficulty 
arises from the fact that the Star Formation Rate (SFR) 
estimation in AGN host galaxies is still problematic. 
All SFR estimation methods present complications and 
even those based on the FIR continuum are
doubtful, since the contribution of the active 
nuclei is unknown (eg. Ho 2005).

Despite the difficulties, some studies, based on different
diagnostics seem to conclude that there is indeed an evolutionary 
sequence from starburst to type 2 and then to type 1 AGN galaxies 
(e.g. Oliva et al. 1999, Krongold et al. 2002). In addition, Kim, Ho and Im (2006), 
using the [OII] emission line as a SFR indicator, reach the conclusion 
that type 2 are the precursors of type 1 quasars supporting the previous claims.
These studies are based on the observed differences between 
different types of AGNs and resemblance of type 2 
objects to starbursts. This raises doubts about the simplest version of the unification 
scheme of AGNs.
It is true that the recent discovery of 10$\mu m$ 
silicate emission in two luminous quasars implies the presence of
dust, but it is not clear yet what is the spatial distribution of this
material (Siebenmorgen et al. 2005). In addition, silicate
emission is not yet detected in other type 1 objects and thus more 
observations are needed to establish the existence of the dusty torus. 

We can summarize all the previous in two statements: (1) the 
starburst-AGN connection is still not well established and (2) the AGN unification 
model, although successful in interpreting many 
observational facts, remains fragile. From our point of view, our BIRG sample 
offers a homogeneous and complete database, which is ideal for a 
statistical study on these issues.

We will discuss our galaxy samples in section \S 2. 
Our data analysis and results will be presented in \S 3, while in 
section \S 4 we will discuss our results and  present our conclusions. 
Due to the fact that all our samples are local, cosmological 
corrections of galaxy distances are negligible. Throughout our 
paper we use $H_{\circ}=100 \; h$ Mpc.

\section{Observations \& Samples}

\subsection{BIRG Galaxies and Control Sample}

The Bright IRAS sample consists of 87 objects with 
redshifts between 0.008 and 0.018 and was compiled 
from the BIRG survey by Soifer et al. (1989) for the 
northern hemisphere and by Sanders et al. (1995) for 
the southern. It includes only high galactic latitude 
objects ($|b|>30^o$) in order to avoid 
extinction and confusion with galactic stars. All objects 
lay in the luminosity range of $10^{10} \; h^{-2} L_\odot \leq L_{FIR} 
\leq 10^{12} \; h^{-2} L_\odot$. This sample is volume 
limited and a $V/V_{max}$ test gives a value of $0.47\pm 0.05$. 
Since the BIRG survey is highly complete, this sample is 
expected to be as well. More details about the sample 
selection are given in Krongold et al. (2002). In addition 
we have refined the Bright IRAS sample by correcting the 
infrared fluxes using ``The IRAS Revised Bright Galaxy Sample" 
by Sanders et al. (2003). Furthermore for interacting galaxies 
we used the corrected fluxes given by Surace, Sanders and Mazzarella (2004).

We also use the control sample, compiled by Krongold et al. (2002) in
such a way as to reproduce the main characteristics, other than the
infrared emission, of the Bright IRAS sample. Specifically, the control
sample was compiled from the original CfA catalog to reproduce closely the
redshift, morphological type and diameter size distributions 
of the corresponding IRAS sample.
In other words, the selection of the IRAS sample and its corresponding
control sample is exactly the same, the only difference being the infrared
luminosity. This is very important in order to validate that any possible
environmental effect is related to the mechanisms that produce the  
observed high infrared luminosity and not to
possible differences in the host galaxies or sample biases. 

In Table 1 we present the names, celestial coordinates, Zwicky
magnitudes, redshifts nearest neighbor projected linear distance 
and spectral types of our final list of Bright IRAS galaxies. 

\subsection{SSRS and CfA2 catalogues}

In order to investigate the local and large scale environment around
our BIRG and control sample galaxies we use the CfA2 and SSRS galaxy
catalogues which cover a large solid angle of the sky. Although these
galaxy catalogues date from the 80's and 90's they still
provide an important database for studies of the properties of
galaxies and their large-scale distribution in the nearby Universe.  
We briefly present the main characteristics of these catalogues.

The CfA2 redshift catalog contains approximately 18000 galaxy
redshifts in the northern sky down to a magnitude limit of
$m_B=$15.5 (Huchra 1990). The magnitude system
used is the merging of the original Zwicky magnitudes and the more
accurate RC1 $B(0)$ magnitudes. These exhibit a scatter of $\sim 0.3$
mags (eg. Bothun \& Cornell 1990). 
Following Huchra (1990), we do not attempt to translate
these magnitudes to a standard photometric system since this 
requires accurate knowledge of the morphological type and size of each
individual galaxy.

The SSRS catalog (da Costa et al. 1998) contains redshifts, 
$B$ magnitudes and morphological classifications for 
$\sim$5400 galaxies
in two regions covering a total of 1.70 steradians in the southern 
celestial hemisphere and it is more than 99\% complete down to $m_{B}$ =
15.5. The galaxies have positions accurate to about 1 arcsec and magnitudes 
with an rms scatter of about 0.3 mag. The radial velocity
precision is of $\sim$ 40 km/s.

Note that in the regions covered by the SSRS and CfA2 catalogues, 
only a subsample of the original BIRGs and their control
samples can be found (76 Bright IRAS galaxies and 61 control 
galaxies). In order to test whether these subsamples are statistically
equivalent with their parent samples (ie., their diameter,
morphological type and redshift distributions) 
we used the Kolmogorov-Smirnov two-sample test. 
We verified that the null hypothesis, the subsamples being
equivalent with their parent samples, cannot be rejected at any
significant statistical level.

\subsection{Our spectroscopic observations}

In order to cover a larger magnitude difference between the BIRGs and
their nearest neighbor than that imposed by the CFA2/SSRS magnitude
limit ($m_B\sim 15.5$) we have obtained our own spectroscopic observations
of fainter neighbors around a subsample of our BIRGs, 
consisting of 24 galaxies (selected randomly from their parent
sample). Around each BIRG we have obtained spectra
of all neighboring galaxies within a projected radius of 100 $h^{-1}$ kpc 
and a magnitude limit of $m_B\lesssim 19.0$. 

Our aim with this new fainter neighbor search 
is not to establish or not the
existence of close neighbors around the BIRGs. This will be done by
using the brighter SSRS and CfA2 catalogues, at the magnitude limit of which
we have well defined control samples.
What we seek with these observations is to facilitate a comparison with
a similar analysis of Seyfert galaxies by Koulouridis et al. (2006),
in which Sy2's were found to have significantly higher fraction
of neighbors with respect to Sy1's. In other words we wish to
establish whether the fractional differences in-between the Sy1,
Sy2 and BIRG samples of galaxies, already determined (or not) 
as significant with respect to their control samples, 
continue to fainter magnitudes.


Optical spectroscopy was carried out using the Faint
Object Spectrograph and Camera (LFOSC) (Zickgraf et al. 1997)
mounted on the 2.1m Guillermo Haro telescope in Cananea,  
operated by the National Institute of Astrophysics, Optics and 
Electronics (INAOE) of Mexico. A setup covering the spectral range 
$4200-9000$\AA\ with a dispersion of 8.2\,\AA/pix was adopted. 
The effective instrumental spectral resolution was about 18\,\AA. 
The data reduction was done using the IRAF packages and included bias and
flat field corrections, cosmic ray cleaning, wavelength linearization,
and flux transformation.

In Table 2 we present the BIRG name, coordinates, redshift and
magnitude for this subsample of BIRGs.
Below the row of each BIRG we list the corresponding data for all its
neighbors, within a projected separation of 100 $h^{-1}$ kpc.
Since Zwicky magnitudes were not available for the fainter neighbors
and in order to provide a homogeneous magnitude system for all the
galaxies we decided to list in Table 2 the $O_{\rm MAPS}$ 
magnitudes\footnote{$O$ (blue) POSS I plate magnitudes of the Minnesota Automated Plate
Scanner (MAPS) system.} for all  galaxies, being the central BIRG or their 
neighbors (see {\tt http://aps.umn.edu/docs/photometry}). The neighbor measured 
redshift is presented in the fifth column (while in some very few
cases we list the redshift from the NED). The uncertainties listed are
estimated from the redshift differences which result from 
using different emission lines.

\section{Analysis and Results}

We search for the nearest neighbor around each BIRG
and control galaxy in our samples with the aim of estimating the fraction
of BIRG and normal galaxies that have a close neighbor. To define the
neighborhood search we use two parameters, 
the projected linear distance ($D$) and the radial velocity        
separation ($\delta v$) between the central BIRG and the neighboring
galaxies found in the CfA2 and SSRS
catalogues or in our own spectroscopic observations.
We search for neighbors with
$\delta v \leq$ 600 km/s, which is roughly the
mean galaxy pairwise velocity of the CfA2 and SSRS galaxies or about 
twice the mean pairwise galaxy velocity when clusters of galaxies are
excluded (Marzke et al. 1995). Note however that our results remain
robust even for $\delta v \leq$ 1000 km/s.
We then define the fraction of BIRG and normal galaxies that
have their nearest neighbor within the selected $\delta v$
separation, as a function of increasing $D$.

\subsection{Neighbors with $m_B \mincir 15.5$ (CfA2 \& SSRS)}

In Figure 1 (upper panels) we plot the fraction of BIRG and control
galaxies as a function of the projected distance ($D$) 
of the first companion and for two velocity separations ($\delta v\le 200$
km/s and $\delta v\le 600$ km/s). For comparison we also plot  
the results of a similar analysis by Koulouridis et al. (2006) 
concerning Seyfert galaxies and their control samples. 

It is evident that a significantly higher fraction of BIRG
galaxies have a near neighbor within $D\mincir 100$ h$^{-1}$ kpc
with respect to their control sample.
In Koulouridis et al. (2006) we found that there is a significantly higher fraction of Sy2
galaxies ($\sim$ 27\%) having a near neighbor within $D\mincir 75$ h$^{-1}$ kpc
with respect to both their control sample and the Sy1 galaxies ($\sim$
14\%).
 Adding here the BIRG sample, which includes mostly starburst and 
Sy2 galaxies (see Table 1), we can clearly see that an even 
higher fraction of BIRGs ($\sim$ 42\%) tend to have a close companion  
within $D\mincir 75$ h$^{-1}$ kpc. The latter needs a further 
explanation since it is not consistent with most starburst-AGN connection 
scenarios, which suggest a simultaneous creation of starburst and 
Sy2 nuclei triggered by interactions.


In order to investigate whether fainter neighbors, than those found in
the relatively shallow CFA2 and SSRS catalogues, exist around our
BIRGs, we have performed a
spectroscopic survey of all neighbors with $m_B \mincir 19.0$ ($\magcir$3
magnitudes fainter than the CfA2 and SSRS limits). This limit
translates into an absolute magnitude limit of M$_B\sim -$15.2 for the
most distant objects in our sample (z$=0.018$). This magnitude is
fainter even than that of the Small Magellanic Cloud. 

\subsection{Neighbors with $m_B \mincir 19.0$ (our spectroscopy)}

Here we present results of our spectroscopic observations 
of all the neighbors with $D\le 75 \; h^{-1}$ kpc and $m_B\mincir 19.0$ for 
the subsample of 24 BIRG galaxies (see Table 2).


We find that in total 13 out of the 24 BIRGs have at least one close
neighbor within the above limits, with 9 of these having
neighbors already detected in the SSRS and CfA2 catalogues, ie., only 4
BIRGs have fainter than $m_b\sim 15.5$ neighbors. This implies that
the BIRGs having a close neighbor (within $D\le 75 \; h^{-1}$ kpc and 
for $\delta v \le 600$ km/s) increases only by $\sim$45\% when going fainter.

Koulouridis et al. (2006) showed that the percentage of 
both Sy1 and Sy2 galaxies that have a close neighbor (within the
above limits) increases correspondingly by about 100\% 
when we descent from $m_B\mincir 15.5$ to $m_B\mincir 19.0$ (but remember that
the host galaxies have magnitudes slightly closer to the CFA2 and SSRS limit). 
In detail, while the percentage of Sy1 and Sy2 galaxies having 
a close neighbor increases from 14\% to 27\% and from 27\% to 55\% 
respectively, for BIRGs it increases from 42\% to 54\%, reaching the
equivalent Sy2 levels. 
We summarize that BIRGs, with respect to their control sample, 
show an excess of close neighbors which therefore should be responsible for 
their excess FIR emission. These results confirm a previous 2-dimensional 
analysis of Krongold et al. (2002) of the same BIRG sample. 

Since the fractions of both BIRGs and Sy2s that have a close neighbor
is roughly the same, an interesting question is whether there are any magnitude
differences between hosts and neighbors for the BIRGs and the Sy2s.
In Figure 2 we present the distribution of such magnitude differences
($\Delta m$) between hosts and nearest neighbor for the BIRGs and the
Sy2s. Although there appears to be a slight preference for brighter
neighbors of the BIRGs with respect to the Sy2s, the two distributions
are statistically equivalent, as quantified by a K-S test which gives a probability of
them being drawn from the same parent population of $\sim 0.75$.



\subsection{Large scale environmental analysis}

Here we investigate whether there are differences in the large scale
environment of BIRGs with respect to their control galaxies and to the
Sy1 and Sy2 samples of Koulouridis et al. (2006).
To this end we determine the galaxy overdensity, based on the CfA2 and 
SSRS catalogues,  in a region around each BIRG or control sample galaxy.
We count all neighboring galaxies around each galaxy within a
projected radius of 1 $h^{-1}$ Mpc, while to
take into account the galaxy peculiar velocities, we use a radial velocity 
separation of $\delta v \le 1000$ km/s.

We estimate the expected CfA2 and SSRS field galaxy density,
$\langle \rho \rangle$,  at the distance of each galaxy by 
integrating the corresponding CfA2 or SSRS luminosity function
(Marzke, Huchra \&
Geller 1994; da Costa et al. 1994) using as a lower integration limit
the minimum galaxy luminosity that corresponds to
the galaxy catalogue magnitude limit (ie., $m_B=15.5$)
at that distance.
We then compute the local overdensity around each AGN, within the
previously mentioned cylinder, which is given by:
$\Delta \rho =(\rho - \langle \rho  \rangle)/\langle \rho \rangle$, 
where $\rho=N/V$ with $N$ the number of neighbors and
$V$ the corresponding volume of the cylinder.

In Figure 3 we plot the overdensity frequency distribution 
for the BIRGs (left panel) and
the corresponding distribution of their control galaxy sample. 
For comparison, we also plot the distributions for the Seyfert
galaxies of Koulouridis et al. (2006). 
A Kolmogorov-Smirnov test shows that there is no statistically
significant difference between any active galaxy sample (BIRG or
Seyfert) and their 
respective control sample distribution. 

However, there is a statistically significant difference, at a 0.03 and 0.09
level, between the overdensity distributions of BIRGs - Sy1s 
and Sy1s - Sy2s, respectively. Similar, differences are also found 
between their respective control samples. On the other hand the
corresponding distributions of BIRGs and Sy2s (and of their control
samples) are statistically equivalent at a 0.9 level.
This implies that the large scale environment of BIRGs and Sy2s is 
similar but significantly different to that of Sy1s, a difference
which since it is also seen in their corresponding control samples
 should be attributed to differences of the host galaxies.
Indeed, Sy1 hosts are earlier type galaxies (e.g., 
Hunt \& Malkan 1999, Koulouridis et al. 2006) which are known to be 
more clustered than late types (e.g., Willmer et al. 1998).

\subsection{FIR color analysis} 

In these section we investigate whether there is relation 
between the strength of the interaction of BIRGs with their
 closest neighbor and their FIR characteristics. The strength
 of any interaction could be parametrized as a function of the distance between 
the BIRG and its first neighbor. At this first order analysis we do not take into 
account the magnitude difference between BIRGs and their close
neighbor, which as we have shown previously (see figure 2), does not
appear to be significantly differ from that of Sy2 galaxies.
We divided the interactions in our sample 
into three categories based on the proximity of the first 
neighbor. We consider strong interactions when 
$D\leq 30 \; h^{-1} \; kpc$, weak interactions when $30\leq D\leq 100
\; h^{-1} \; kpc$ and no interaction when $D> 100 \; h^{-1} \; kpc$.  

In Figure 4 we present the color - color diagram of $\alpha(60,25)$ 
versus $\alpha(25,12)$, where $\alpha(\lambda_1, \lambda_2)$ is 
the spectral index defined as 
$\alpha(\lambda_1,
\lambda_2)=log(S_{\lambda_1}/S_{\lambda_2})/(\lambda_2/\lambda_1)$ 
with $S_{\lambda_1}$ the flux in janskys at wavelength $\lambda_1$. 
We can clearly see the differences between the FIR characteristics 
among different types of galaxies and different interaction strengths.
The different interaction strengths are coded by different
types of symbols while the different activity is color coded as indicated 
in the caption of the figure. We cross-identified classifications 
of each BIRG combining various studies like : The Pico Dos Dias Survey 
(Coziol et al. 1998), Optical Spectroscopy of luminous infrared 
galaxies (Veilleux et al. 1997; Ho, Filippenko, Sargent 1995), COLA -- Radio and 
spectroscopic diagnostics of nuclear activity in galaxies (Corbett et al.
2002), Warm Iras Sources (de Grijp et al. 1987) and, when available, SDSS 
spectroscopic data.


It is evident that the FIR characteristics of starburst galaxies in 
our BIRG sample differ significantly depending on the strength of the interaction. 
The majority of highly interacting starburst have $\alpha(60,25)$ spectral 
indices greater than -2, while all, except one, non-interacting starbursts have less. 
We also find that normal galaxies and Liners are at the lower end of this sequence. 
The only highly-interacting starburst galaxy below $\alpha(60,25)=-2.15$ 
is NGC7541 which happens to have more than two times the typical
molecular gas mass of BIRGs (Mirabel \& Sanders 1988). The difference
between highly interacting and non-interacting starburst galaxies, 
as quantified by a K-S test, is significant at a $>99.9\%$ level
when comparing their $\alpha(60,25)$ index distribution. 
However, Sy2 galaxies, interacting or no, 
seem to lay in the same area ($-2.5<\alpha(60,25)<-2$)
with non-interacting starburst galaxies, 
delineated in figure 4 by blue dashed lines.

The FIR color analysis of our sample strengthens our previous results. It 
clearly shows that the starburst activity is higher when
interactions are stronger and ceases when the interacting neighboring 
galaxy moves away. While the starburst activity weakens (if we link position on
the plot with time) Sy2 nuclei appear, giving further evidence on the causal
bridging between these objects.

\section{Discussion \& Conclusions}

We have compared the 3-dimensional environment of a sample of 
local BIRGs with that of a well defined control sample,
selected in such a way as to reproduce the redshift, morphological
type and diameter size distributions of the BIRG sample. 
We searched for close neighbors around each BIRG and control sample galaxy
using the distribution of CfA2 and SSRS galaxy catalogues as well as
our own spectroscopic observations reaching a fainter magnitude
limit (but for a restricted BIRG subsample).
We also compared our results with those of a similar 
analysis of Seyfert galaxies (Dultzin-Hacyan et al. 1999; Koulouridis et al. 2006).

We find that the fraction of BIRG galaxies having a close
neighbor, within a projected separation of 75 $h^{-1}$ kpc
and radial velocity difference of $\delta v \le 600$ km/s, 
is significantly higher than the corresponding fraction
of its control sample and that of Sy1 galaxies while it
is comparable to that of Sy2 galaxies.
This result is in accordance with some previous two-dimensional 
studies (eg. Krongold et al. 2002).
We reach similar results regarding the large scale environment of BIRGs
(within a projected radial separation of 1 $h^{-1}$ Mpc and a radial velocity 
difference $\delta v \le 1000$ km/s). Once more 
their behavior resembles that of Sy2's but not of Sy1's.  
 We also find a statistically significant difference between 
highly-interacting and non-interacting BIRGs, based on their FIR color 
properties. Sy2s 
appear to display a similar behavior to that of non-interacting 
starburst galaxies, introducing new evidence for the 
starburst/AGN connection scenarios. 

Our results can be accommodated in a simple 
evolutionary scenario, starting with an interaction, and ending in a Sy1 phase.
First, close interactions would drive molecular clouds 
towards the central area, creating a circumnuclear starburst. Then, material
could fall even further into the innermost regions of the galaxy, feeding the
black hole, and giving birth to an AGN which at first cannot be observed due to
obscuration. At this stage only a starburst would be observed. As starburst
activity relaxes and obscuration decreases, a Sy2 nucleus would be revealed
(still obscured by the molecular clouds from all viewing angles). As a
final stage, a Sy1 phase could appear. In this case, the molecular clouds, initially in a
spheroidal distribution, could flatten and form a ``torus'' (as in the
unification scheme for Seyferts). As more material is accreted, it is possible
that the AGN strengthens driving away most 
of the obscuring clouds, and leaving a ``naked'' Sy1 nuclei. 

If indeed interactions play a role in triggering activity, as suggested by the
above picture, then the lack of close companions among Sy1 galaxies implies
that the time needed for type 1 activity to
appear should be larger than the timescale for an unbound
companion to escape from the close environment, or comparable to the timescale
needed for an evolved merger ($\sim10^9$ years, see Krongold et al. 2002).  

It should be noted that the evolutionary scenario does not
contradict the unification scheme. It implies that Sy1s and Sy2s
are the same objects (as the unification model proposes) but not necessarily
at the same evolutionary phase. However, there could be a phase where
only orientation could define if an object appears as a Sy1 or as a Sy2,
which is the stage where molecular clouds form a torus but have not been
swept away yet. 

Evidently, more detailed observations of large 
samples of galaxies are needed to resolve this important issue. 
However, such a picture is consistent with the evolutionary scenario suggested
by Tran (2003). He studied a sample of Sy2 in polarized light, and found that
only 50\% of them showed the presence of a hidden broad line
region (HBLR). He suggested that non-HBLR Sy2 galaxies could evolve into 
HBLR Sy2 galaxies. In this case, the appearance of the BLR could be related to
the accretion rate (e.g. Nicastro 2000), and thus with the evolutionary stage
of the object. The trend is also consistent with the finding that 50\% of Sy2
galaxies also show the presence of a strong starburst (Gu et al. 2001;
Cid-Fernandez et al. 2001).

On the brightest end, there is also growing evidence showing 
(a) that ULIRGs can be the precursors of
quasars and (b) ULIRGS are found in very strong
interacting systems or in mergers (e.g. Sanders, Surace and Ishida
1999; Wang et al. 2006). Therefore, the evolutionary sequence proposed
above could be
general to all nuclear activity and independent on luminosity (we note that
Krongold et al. 2003 suggested a similar scheme for LINERs that can be
considered as the very low luminosity extension of the evolutionary
model suggested here). Further evidence comes from the fact that type
2 quasars also tend to be in interaction more often than type 1
quasars (Serber et al. 2006).

In order to better understand the role of interactions in driving starburst and 
nuclear activity (and the validity of the evolutionary trend), we are in  the 
process of studying AGN and starburst manifestations in  the nearest neighbors 
of the active galaxies in our samples, since the same physical processes should 
act on both members of the pair (host and nearest neighbor).

\acknowledgments

EK thanks the IA--UNAM for its warm hospitality were a major part of this work
was completed. MP acknowledges funding by the Mexican Government research grant
No. CONACyT 39679, VC by the CONACyT research grants 39560-F, 42609 and D. D-H 
support from grant IN100703 from DGAPA, PAPIIT, UNAM. This research has made use 
of the MAPS Catalog of POSS I supported by the University of Minnesota (the APS 
databases can be accessed at {\tt http://aps.umn.edu/}) and of the USNOFS Image 
and Catalogue Archive operated by the United States Naval Observatory, Flagstaff 
Station ({\tt http://www.nofs.navy.mil/ data/fchpix/}).

\begin{deluxetable}{lrrcccccc}
\tabletypesize{\scriptsize}
\tablecaption{Our Bright IRAS sample galaxies which reside in the sky region covered
  by the SSRS and Cfa2 catalogues.}
\tablewidth{0pt}
\tablehead{
\colhead{NAME} & \colhead{RA (J2000)} & \colhead{DEC (J2000)} 
& \colhead{$m_B$} & \colhead{z} &\colhead{D (h$^{-1}$ kpc)} &\colhead{TYPE}  
}
\startdata

NGC0023      &    00 09  53.1 & 25  55  25 &12.50 &0.0152 & isolated &starburst\\
ESO079-G003   &   00 14   54.7& $-$39 11  19 &12.50 &0.0090 & isolated &normal\\
NGC0174      &    00 36   59.0& $-$29  28  40 &13.62 &0.0116 & isolated &starburst\\
UGC00556     &    00 54   49.6 & 29  14  43 &15.30 &0.0154 & isolated & Liner\\
UGC00903     &    01 21   47.1 & 17  35  34 &14.70 &0.0084 & isolated & unclassified\\
ESO353-G020  &    01 34   51.6& $-$36   08   08 &13.95 &0.0161 & isolated &normal\\
NGC0716      &    01 52   59.3 & 12  42  31 &14.00 &0.0152 & isolated &normal\\
UGC01451     &    01 58   29.9 & 25  21  34 &14.30 &0.0164 & isolated &normal\\
NGC0835      &    02 09   24.5& $-$10   08   06 &13.14 &0.0138 &11.00  &starburst\\
NGC0838      &    02 09   38.3& $-$10   08  45 &14.22 &0.0128 &27.82  &starburst\\
NGC0839      &    02 09   42.8& $-$10  10  59 &14.20 &0.0128 &27.91  &starburst\\
NGC0873      &    02 16   32.2& $-$11  20  52 &12.83 &0.0134 &isolated &starburst\\
NGC0877      &    02 17   58.5 & 14  32  53 &12.50 &0.0131 &isolated  &starburst\\
NGC0922      &    02 25    04.4& $-$24  47  15 &12.63 &0.0103 &isolated &starburst\\
NGC0992      &    02 37   25.2 & 21   06   06 &13.50 &0.0138 & 30.32  &starburst\\
NGC1083       &   02 45   40.5& $-$15  21  24 &15.19 &0.0137 &isolated &starburst\\
NGC1134        &  02 53   40.9 & 13  00 58 &13.20 &0.0121 &77.18 &normal\\
UGC02403       &  02 55   57.2  &00 41  36 &13.20 &0.0139 &isolated &starburst\\
NGC1204        &  03 04   39.9& $-$12  20  25 &14.21 &0.0143 & isolated &starburst\\
ESO420-G013    &  04 13   49.5& $-$32  00 23 &13.52 &0.0121 & isolated & Sy2\\
NGC1614        &  04 31   35.5& $-$08  40  56 &12.00 &0.0160 & 2.00 & starburst\\
NGC2782        &  09 14    05.5 & 40   06  54 &12.66 &0.0085 & isolated &starburst\\
NGC1667        &  04 46   10.5& $-$06  24  24 &13.00 &0.0150 & isolated & Sy2\\
NGC2785        &  09 15   15.8 & 40  55   08 &14.90 &0.0088 &52.00  &starburst\\
NGC2856        &  09 24   16.9 & 49  14  58 &13.90 &0.0088 & 26.53  &starburst\\
NGC3147        &  10 16   55.8 & 73  24   07 &11.52 &0.0094 &isolated & Sy2\\
NGC3221        &  10 22   21.0 & 21  34  12 &14.30 &0.0137 & isolated &normal\\
NGC3367        &  10 46   34.5 & 13  45  10 &12.22 & 0.0101 & isolated & unclassified\\
NGC3508        &  11 02   59.6& $-$16  17  17 &13.20 &0.0130 & isolated &starburst\\
*NGC3690       &  11 28   32.9 & 58  33  19 &13.20 &0.0104 & 2.16  &starburst\\
NGC3735        &  11 36    01.0 & 70  32   06 &12.60 &0.0090 &isolated  &Sy2\\
NGC3994        &  11 57   37.0 & 32  16  39 &13.68 &0.0105 & 18.17  &Liner\\
NGC3995        &  11 57   44.9 & 32  17  42 &12.96 &0.0109 & 18.65  &starburst\\
NGC4175        &  12 12   30.7 & 29  10  10 &14.20 &0.0131 & 17.39  &Sy2\\
NGC4194        &  12 14   10.1 & 54  31  34 &13.00 &0.0084 & isolated  &starburst\\
NGC4332        &  12 22   47.8 & 65  50  36 &13.20 &0.0091 & isolated &starburst\\
NGC4388        &  12 25   46.7 & 12  39  40 &12.20 &0.0084 &isolated  &Sy2\\
NGC4433        &  12 27   38.6 & $-$08  16  49 &12.90 &0.0099 & 63.09 &normal\\
MCG-02-33-098  &  13 02   19.8& $-$15  46   07 &15.38 &0.0159 & isolated  &starburst\\
MCG-03-34-014  &  13 12   34.5& $-$17  32  28 &13.02 &0.0092 & isolated &normal\\
NGC5020        &  13 12   40.1 & 12  35  57 &13.40 &0.0112 & isolated &normal\\
IC0860         &  13  15   03.7 &  24  37  05 & 14.80 & 0.0129 & isolated & unclassified \\
NGC5073        &  13 19   21.4& $-$14  51  46 &13.50 &0.0091 & isolated &starburst\\
IC4280         &  13 32   53.0& $-$24  12  29 &13.51 &0.0163 & 93.52  &starburst\\
NGC5371        &  13 55   40.3 & 40  27  38 &11.59 &0.0085 & isolated  &unclassified\\
NGC5394        &  13 58   33.7 & 37  27  10 &13.85 &0.0116 & 19.44  &starburst\\
NGC5395        &  13 58   37.9 & 37  25  27 &12.47 &0.0116 & 19.10  &Liner\\
NGC5430        &  14 00  45.7 & 59  19  46 &13.08 &0.0099 & isolated &starburst\\
NGC5433        &  14 02   36.1 & 32  30  35 &14.00 &0.0145 & isolated &starburst\\
NGC5427        &  14 03   25.7 & $-$06   01  53 &11.93 &0.0087 & 22.81  &Sy2\\
NGC5595        &  14 24   13.2& $-$16  43  28 &13.12 &0.0090 &  32.63 &normal\\
NGC5597        &  14 24   27.2& $-$16  45  50 &13.32 &0.0089 & 32.68  &starburst\\
NGC5653        &  14 30   10.3 & 31  12  50 &13.39 &0.0119 &  isolated &starburst\\
NGC5728        &  14 42   23.6& $-$17  15  14 &12.81 &0.0095 &isolated  &Sy2\\
NGC5757        &  14 47   46.2& $-$19   04  45 &13.50 &0.0089 &isolated &starburst\\
NGC5793        &  14 59   24.6& $-$16  41  38 &14.17 &0.0117 &37.36  &Sy2\\
UGC09668       &  14 55   56.0 & 83  31  29 &13.80 &0.0131 & 63.55  &starburst\\
CGCG049-057    &  15 13   13.2 &  07  13  26 &15.50 &0.0130 & isolated  &starburst\\
NGC5900        &  15 15    05.0 & 42  12  28 &15.00 &0.0084 & 70.48 &normal\\
NGC5930        &  15 26    07.8 & 41  40  30 &13.00 &0.0087 & 3.59  &starburst\\
NGC5936        &  15 30    01.1 & 12  59  21 &13.41 &0.0134 &  isolated  &starburst\\
NGC5937        &  15 30   46.0 & $-$02  49  49 &13.35 &0.0095 &isolated  &starburst\\
NGC5990        &  15 46   16.0 &  02  24  49 &13.10 &0.0128 & 84.90  &Sy2\\
*NGC6052       &  16 05   13.1 & 20  32  27 &14.70 &0.0151 &  isolated  &starburst\\
ESO402-G026    &  21 22   31.7& $-$36  40  57 &13.69 &0.0093 &isolated &normal\\
NGC7130        &  21 48   19.5& $-$34  57  10 &13.33 &0.0161 & isolated  &unclassified\\
NGC7172        &  22 02    02.2& $-$31  52  15 &12.95 &0.0086 & 46.75 & Sy2\\
IC5179         &  22 16    09.3& $-$36  50  43 &12.46 &0.0114 &isolated & starburst\\
ESO534-G009    &  22 38   41.7& $-$25  51   02 &13.55 &0.0113 &  86.40  &Liner\\
NGC7469        &  23 03   15.5 & 08  52  24 &13.00 &0.0162 &  18.76  &unclassified\\
NGC7541        &  23 14   43.0 &  04  32   03 &12.70 &0.0089 & 23.62  &starburst\\
NGC7591        &  23 18   16.0 &  06  35   08 &13.80 &0.0165 &isolated  &Sy2\\
NGC7678        &  23 28   27.7 & 22  25  15 &12.70 &0.0116 & isolated  &Sy2\\
NGC7714        &  23 36   14.0 &  02   09  17 &13.10 &0.0093 & 16.12  &starburst\\
NGC7769        &  23 51    04.7 & 20   09   03 &13.10 &0.0141 &   2.64  &unclassified\\
NGC7771        &  23 51   24.7 & 20   06  41 &13.39 &0.0143 &  13.23  &unclassified\\
UGC12914       &  24 01   38.0 & 23  29   04 &13.20 &0.0146 &  14.82  &unclassified\\
UGC12195       &  24 01   42.2 & 23  29  41 &13.90 &0.0145 &  14.33  &unclassified\\

\enddata
\end{deluxetable}
\clearpage

\begin{deluxetable}{lrrccc}
\tabletypesize{\scriptsize}
\tablecaption{The subsample of BIRG galaxies in our spectroscopic
  survey. Below each BIRG we list all their neighbors within a
  projected distance of 100 $h^{-1}$ kpc with their measured redshifts.}

\tablewidth{0pt}
\tablehead{
\colhead{NAME}         &\colhead{RA}       &\colhead{DEC}     &\colhead{$O_{\rm MAPS}$} & \colhead{z}            & \colhead{TYPE} \\
                       &\colhead{J2000}  &\colhead{J2000} & \colhead{integrated} &                           &
}
\startdata
NGC0023      &    00 09  53.1 & 25  55  25 & 13.95   &0.0152 & starburst \\
\,\,\,\,none\\
UGC00556     &    00 54   49.6 & 29  14  43 & 15.63&0.0154 & Liner\\
\,\,\,\,neighbor 1 & 00 54 51.1  &  29 16 25&17.03  & 0.0152$\pm$0.0002 \\
NGC0716      &    01 52   59.3 & 12  42  31 &14.51 &0.0152  &Normal\\
\,\,\,\,none\\
UGC01451     &    01 58   29.9 & 25  21  34 &15.41 &0.0164 &Normal\\
\,\,\,\,none\\
NGC0835      &    02 09   24.5& $-$10   08   06 &13.67$\star$ &0.0138 &starburst\\
\,\,\,\,neighbor 1 & 02 09 20.9 & $-$10 08 00 &$\star$ &0.0130$\pm$0.0002\\
\,\,\,\,neighbor 2 & 02 09   38.3& $-$10  08  45 &14.89 &0.0133$\pm$0.0004 \\
\,\,\,\,neighbor 3 & 02 09   42.8& $-$10  10  59 &15.01 &0.0132$\pm$0.0004 \\
NGC0877      &    02 17   58.5 & 14  32  53 & 13.07  &0.0131 &starburst\\
\,\,\,\,neighbor 1 & 02 17 53.3  &  14 31 17 &16.04 & 0.0136$\pm$ 0.0007\\
\,\,\,\,neighbor 2 & 02 17 26.3 &  14 34 49 &16.77 & 0.013376$\dagger\dagger$\\
NGC0922      &    02 25    04.4& $-$24  47  15 &13.25 &0.0103 &starburst\\
\,\,\,\,neighbor 1 & 02 24 30.0 & $-$24 44 44 & 16.73&0.1054$\dagger\dagger$\\
NGC0992      &    02 37   25.2 & 21   06   06 &15.39 &0.0138   &starburst\\
\,\,\,\,neighbor 1 & 02 37 28.2 & 21 08 31 &16.99 & 0.0126$\pm$ 0.0004\\
NGC1614        &  04 31   35.5& $-$08  40  56 &14.55   &0.0160  & Sy2\\
\,\,\,\,neighbor 1 &  04 31   35.5& $-$08  40  56 & 16.44  &0.0160\\
NGC1667        &  04 46   10.5& $-$06  24  24 &13.00$\dagger$ &0.0160  & Sy2\\
\,\,\,\,none\\
UGC02403       &  02 55   57.2  &00 41  36 &15.33   &0.0139 &starburst\\
\,\,\,\,neighbor 1 & 02 55 58.9 & 00 40 26 &19.10   &0.0749$\pm$ 0.0002\\
NGC2785        &  09 15   15.8 & 40  55   08 &14.85 &0.0088  &starburst\\
\,\,\,\,neighbor 1 & 09 15 33.8 & 40 55 27 &14.54 & 0.0653$\pm$0.0004\\
\,\,\,\,neighbor 2 & 09 14 43.1 & 40 52 47 &14.54 & 0.008319$\dagger\dagger$\\
\,\,\,\,neighbor 3 & 09 14 35.6 & 40 55 24 & 17.58  & 0.008933$\dagger\dagger$\\
NGC2856        &  09 24   16.9 & 49  14  58 &14.71 &0.0088  &starburst\\
\,\,\,\,neighbor 1 & 09 24 03.1 & 49 12 16 &14.52 & 0.0089$\pm$0.0004\\
NGC3221        &  10 22   21.0 & 21  34  12 &13.87   &0.0137 &Normal\\
\,\,\,\,neighbor 1 & 10 22 26.0 & 21 32 31 &17.06 & 0.0117$\pm$0.0004\\
\,\,\,\,neighbor 2 & 10 22 21.1 & 21 31 00 &17.89 & 0.0539$\pm$0.0003\\
\,\,\,\,neighbor 3 & 10 22 13.4 & 21 30 42 &18.67& 0.0128$\pm$0.0007\\
NGC3690       &  11 28   32.9 & 58  33  19 &13.76$\star$ &0.0104  &starburst\\
\,\,\,\,neighbor 1 & 11 28 33.5 & 58 33 47 &$\star$ & 0.010411$\dagger\dagger$\\
\,\,\,\,neighbor 2 & 11 28 27.3 & 58 34 42 &$\star$ & 0.0132$\pm$0.0001\\
\,\,\,\,neighbor 3 & 11 28 45.8 & 58 35 36 & 16.14  & 0.0604$\pm$0.0002 \\
NGC4388        &  12 25   46.7 & 12  39  40 & 12.79&0.0084   &Sy2\\
\,\,\,\,neighbor 1 & 12 25 41.7 & 12 48 38 &14.50 & 0.0021$\pm$0.0001\\
\,\,\,\,neighbor 2 & 12 25 15.2 & 12 42 53 &15.87& $<$0.001$\dagger\dagger$\\
IC0860       &  13  15   03.7 &  24  37  05 &15.31& 0.0129  & unclassified \\
\,\,\,\,none\\
NGC5073        &  13 19   21.4& $-$14  51  46 &14.03   &0.0091 &starburst\\
\,\,\,\,neighbor 1 & 13 19 34.1 & $-$14 46 22 &16.21 & 0.0350$\pm$0.0006\\ 
\,\,\,\,neighbor 2 & 13 18 56.4 & $-$14 54 13 &16.41   & 0.0347$\pm$0.0001\\
NGC5433   &  14 02   36.1 & 32  30  35 &14.68 &0.0145 &starburst\\
\,\,\,\,neighbor 1 &  14 02 39.0 & 32 27 50 &18.00 & 0.0142$\pm$0.0008\\
\,\,\,\,neighbor 2 & 14 02 20.5 & 32 26 53 &16.17 & 0.00141$\pm$0.0007\\
NGC5653        &  14 30   10.3 & 31  12  50 &14.10 &0.0119  &starburst\\
\,\,\,\,none\\
NGC5990        &  15 46   16.0 &  02  24  49 &14.29 &0.0128  &starburst\\
\,\,\,\,neighbor 1 & 15 46 28.9 & 02 23 09 &18.16 & 0.0468$\pm$0.0002\\
\,\,\,\,neighbor 2 & 15 46 23.2 & 02 21 34 &17.58 & 0.0480$\pm$0.0003\\
\,\,\,\,neighbor 3 & 15 45 45.9 & 02 24 35 &15.87 & 0.0141$\pm$0.0001\\
NGC7541   &  23 14   43.0 &  04  32   03 &13.22 &0.0089  &starburst\\
\,\,\,\,neighbor 1 & 23 14 34.5 & 04 29 54 &14.70 & 0.0080$\pm$0.0006\\
NGC7714        &  23 36   14.0 &  02   09  17 &13.10$\dagger$ &0.0093  &starburst\\
\,\,\,\,neighbor 1 & 23 36 22.1 & 02 09 24 &14.90$\dagger$ & 0.0089$\pm$0.0001\\
NGC7771        &  23 51   24.7 & 20   06  41 &13.81$\star$   &0.0143   &unclassified\\
\,\,\,\,neighbor 1 & 23 51 22.5 & 20 05 47 &$\star$ & 0.0145$\pm$0.0008\\
\,\,\,\,neighbor 2 & 23 51 13.1 & 20 06 12 &17.13 & 0.013679$\dagger\dagger$\\
\,\,\,\,neighbor 3 & 23 51 04.0 & 20 09 02 &14.05 & 0.0139$\pm$0.0003\\
\,\,\,\,neighbor 4 & 23 51 13.9 & 20 13 46 &17.02 & 0.043527$\dagger\dagger$\\
\enddata

\tablenotetext{$\dagger$}{Zwicky blue magnitude (Region Not Covered by MAPS Catalog)}
\tablenotetext{$\dagger\dagger$}{Redshift from NED}
\tablenotetext{$\star$}{Not resolved neighboring galaxies}
\end{deluxetable}

\newpage

\begin{figure}
\epsscale{0.8}
\plotone{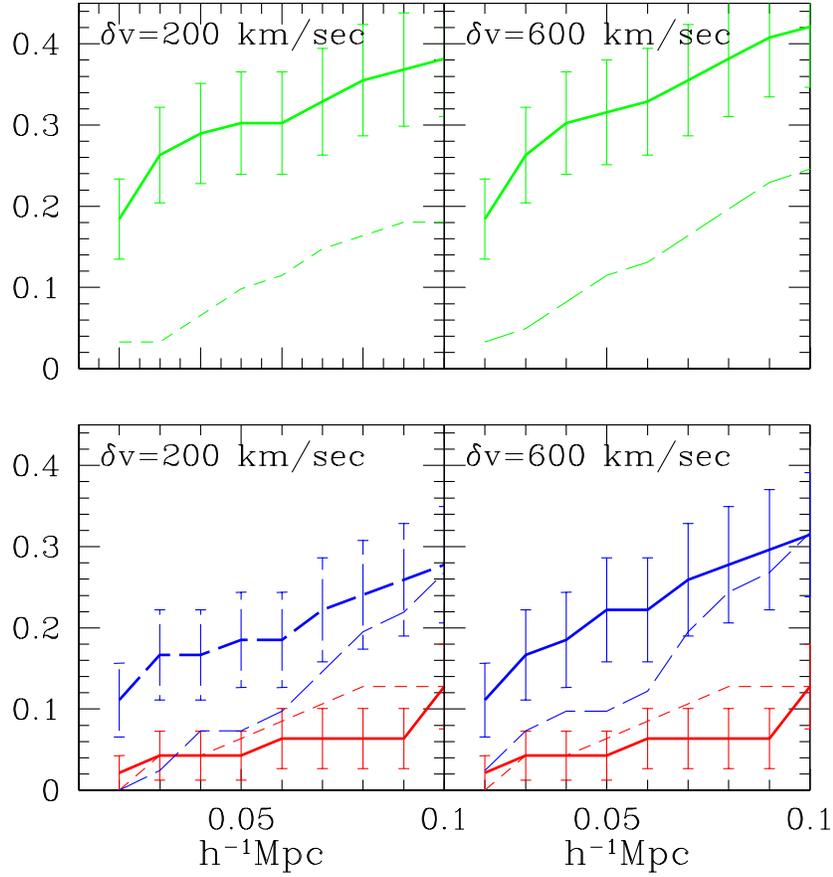}
\figcaption{{\em Top Panels}: Fraction of BIRGs (thick line) 
and their control sample galaxies (thin dashed line) which 
have their nearest neighbor, within the indicated redshift separation,
as a function of projected distance. {\em Bottom Panels}:
 Corresponding Sy1 (red) and Sy2 (blue) fractions by Koulouridis et al. 2006}
\end{figure}

\begin{figure}
\epsscale{0.8}
\plotone{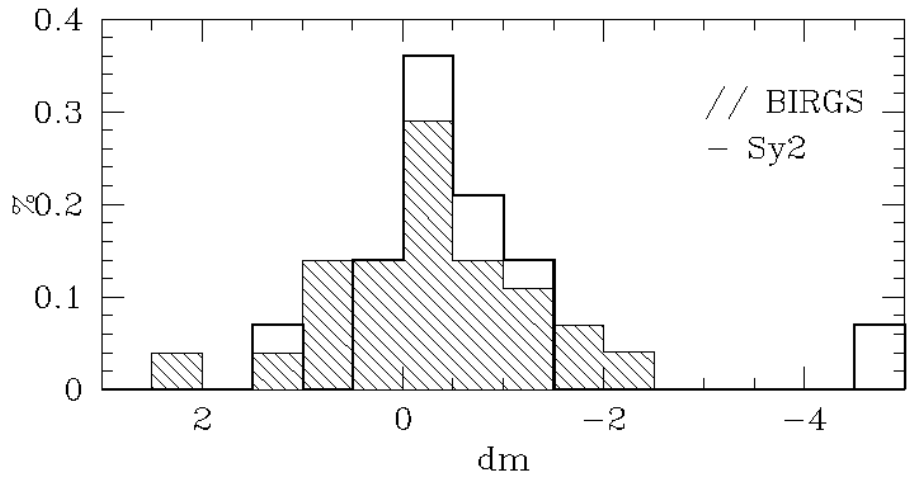}
\figcaption{The frequency distribution of host (BIRG or Sy2) - nearest
neighbor magnitude differences.}
\end{figure}

\newpage

\begin{figure}
\epsscale{0.8}
\plotone{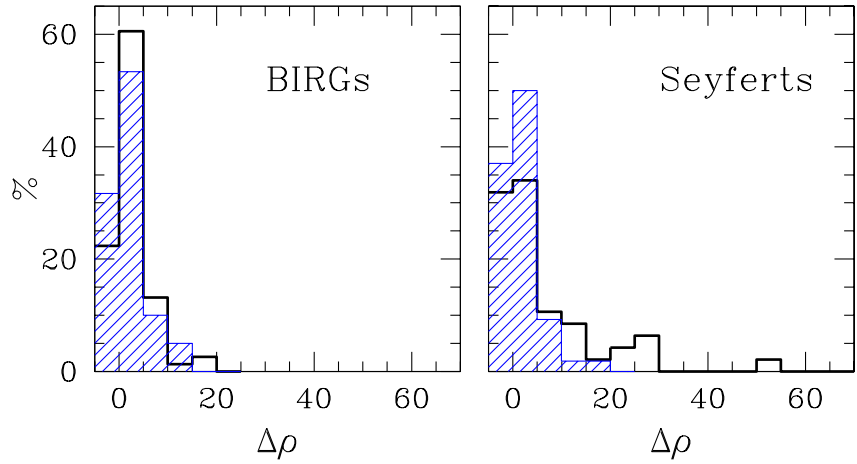}
\figcaption{{\em Left Panel:} Frequency distribution of galaxy overdensities 
around BIRGs (solid line) and control sample (shaded
histogram). {\em Right Panel:}  Corresponding distribution around Sy1s
(solid line) and Sy2s (dashed histogram). 
Note that here we do not
present the corresponding distributions of their control samples (to
this end see Koulouridis et al. 2006).
}
\end{figure}

\begin{figure}
\epsscale{0.8}
\plotone{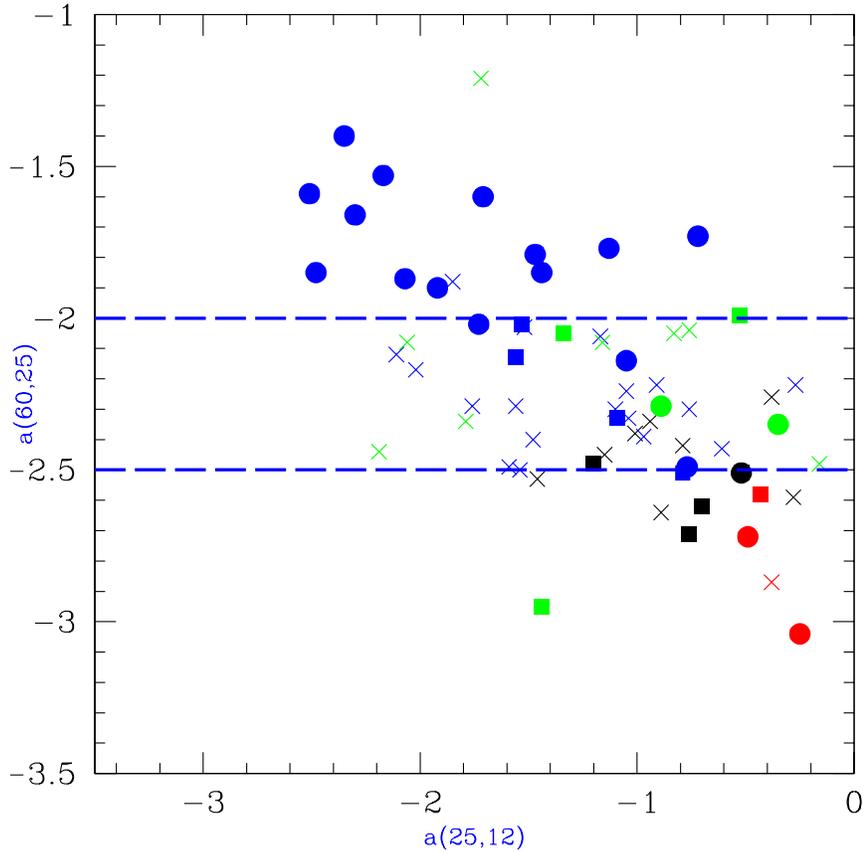}
\figcaption{FIR color-color diagram: $\alpha(60,25)$ versus
  $\alpha(25,12)$. The color coding is such that starbursts are
  represented by blue, Sy2s by green, Liners by red and normal
  galaxies by black. Highly interacting BIRGs are represented 
by filled circles, weakly interacting by filled squares 
and non-interacting by crosses.}
\end{figure}

\end{document}